\documentclass[a4paper,12pt]{article}
\pdfoutput=1

\makeatletter

\@addtoreset{equation}{section}
\makeatother
\usepackage{amsfonts}
\usepackage{amsmath}
\usepackage{latexsym}

\def\be{\begin{equation}}
\def\ee{\end{equation}}
\def\bea{\begin{eqnarray}}
\def\eea{\end{eqnarray}}
\def\({\left(}
\def\){\right)}
\def\<{\left<}
\def\>{\right>}

\def\[{\left[}
\def\]{\right]}

\def\be{\begin{equation}}
\def\ee{\end{equation}}
\def\bea{\begin{eqnarray}}
\def\eea{\end{eqnarray}}

\def\({\left(}
\def\){\right)}
\def\<{\left<}
\def\>{\right>}

\def\tr{{\mbox{tr}}}
\def\be{\begin{equation}}
\def\ee{\end{equation}}
\def\bea{\begin{eqnarray*}}
\def\eea{\end{eqnarray*}}
\def\ben{\begin{eqnarray}}
\def\een{\end{eqnarray}}
\def\({\left(}
\def\){\right)}
\def\<{\left<}
\def\>{\right>}
\def\!{\right|}
\def\|{\left|}

\def\[{\left[}
\def\]{\right]}

\def\+{\bar}
\def\mb{\mathbb}
\def\tr{{\mbox{tr}}}

\def\Vol{{\mbox{Vol}}}

\def\L{{\cal{L}}}
\def\t{\widetilde}
\def\A{{\cal{A}}}

\def\M{{\cal{M}}}

\def\N{{\cal{N}}}

\def\O{{\cal{O}}}

\def\L{{\cal{L}}}

\def\L{{\cal{L}}}

\def\l{{{\ell}}}

\def\h{\widehat}

\begin{document}

\pagestyle{empty}
\vskip-10pt
\vskip-10pt
\begin{center}
\vskip 3truecm
{\Large\bf
Abelian M5-brane on $S^6$}
\vskip 2truecm
{\large \bf
Andreas Gustavsson}
\vspace{1cm} 
\begin{center} 
Physics Department, University of Seoul, 13 Siripdae, Seoul 130-743 Korea
\end{center}
\vskip 0.7truecm
\begin{center}
(\tt agbrev@gmail.com)
\end{center}
\end{center}
\vskip 2truecm
{\abstract{We study the abelian M5 brane on $S^6$. From the spectrum we extract a series expansion for the heat kernel. In particular we determine the normalization for the  coefficient $a$ in the M5 brane conformal anomaly. When we compare our result with what one gets by computing the Hadamard-Minakshisundaram-DeWitt-Seeley coefficients from local curvature invariants on $S^6$, we first find a mismatch of one unit. This mismatch is due to an overcounting of one zero mode. After subtracting this contribution, we finally find agreement. We perform dimensional reduction along a singular circle fiber to five dimensions where we find the conformal anomaly vanishes.}}

\vfill
\vskip4pt
\eject
\pagestyle{plain}

\section{Introduction}
The M5 brane conformal anomaly was computed on the gravity side in \cite{Henningson:1998gx} and for the abelian M5 brane in \cite{Bastianelli:2000hi} by extracting it from the Hadamard-Minakshisundaram-DeWitt-Seeley (HMDS) coefficient $a_6$ in the heat kernel expansion. The coefficient $a_6$ was expressed in terms of $46$ local invariants by Gilkey \cite{Gilkey:1975iq} for a smooth compact Riemannian six-manifold $M$ with metric $g_{\mu\nu}$ and for a second order, elliptic, positive definite differential operator of the form
\bea
D &=& - g^{\mu\nu} D_{\mu} D_{\nu} - E
\eea
If there is a gauge bundle over $M$, then $E$ will be matrix valued in that gauge bundle and $D_{\mu}$ will involve both the Christoffel symbol as well as the gauge bundle connection. We follow the notation of the review paper \cite{Vassilevich:2003xt}. The M5 brane conformal anomaly has the general form
\ben
\A &=& a E_6 + c_1 I_1 + c_2 I_2 + c_3 I_3 + D_i J^i\label{overall}
\een
where $E_6$ is proportional to the Euler density, $I_i$ are a conformal invariants that are constructed out of the Weyl tensor, and $D_i J^i$ is a scheme-dependent total derivative. On the supergravity side the result is \cite{Henningson:1998gx}
\ben
\A &=& \frac{4N^3}{(4\pi)^3 7!} \(-\frac{35}{2} E_6 - 1680 I_1 - 420 I_2 + 140 I_3 + D_i J^i\)\label{N}
\een
For the abelian M5 brane, the result that one gets by applying Gilkey's formula for $a_6$ is \cite{Bastianelli:2000hi} 
\bea
\A &=& \frac{1}{(4\pi)^3 7!} \(-\frac{245}{8} E_6 - 1680 I_1 - 420 I_2 + 140 I_3 + D_i J^i\)
\eea
We notice that the $c_i$-coefficients agree up to an overall factor of $4N^3$, while for the coefficient $a$ we would need to add $105/8$ in order to get the same sort of agreement,
\ben
-\frac{245}{8} + \frac{105}{8} &=& -\frac{35}{2}\label{diff}
\een
There is however no reason to expect such an agreement for the $a$-coefficient, as was explained in \cite{Bastianelli:2000hi}. Given the match of the $c_i$-coefficients together with the motivation in \cite{Bastianelli:2000hi} for why such a match should be anticipated, there seems to be little doubt about the correctness of the result for the $c_i$ coefficients for the abelian theory. But there is no such corresponding match for the $a$-coefficient, nor has there been any independent computation of the $a$-coefficient in the  literature. Therefore we think that it can be motivated to present an independent computation of the $a$-coefficient. Only the combination $a E_6$ has an invariant significance, but not $a$ in isolation since we can always rescale $E_6$ such that $a=1$. The result we get for the integrated anomaly on $S^6$ in a first computation is
\bea
\int_{S^6} \A = \frac{2}{105} \cdot \frac{245}{8} - 1
\eea
But by a careful examination of zero modes, we trace $-1$ to a zero mode that has been overaccounted for \cite{Christensen:1979iy}, \cite{Fradkin:1983mq}, \cite{Tseytlin:2013fca} and our final result is therefore
\bea
\int_{S^6} \A = \frac{2}{105} \cdot \frac{245}{8}
\eea
in agreement with \cite{Bastianelli:2000hi}. There are many indirect evidences that suggest that this value for $a E_6$ is the correct one \cite{M1}, \cite{M2}, \cite{M3}. Since $S^6$ is conformally flat, the Weyl tensor is zero and so $I_i=0$. The only term that survives in the integrated conformal anomaly on $S^6$ is the term that is proportional to the Euler characteristic $E_6$. 

By taking into account the normalization for abelian gauge group above, we may then from the result in \cite{Maxfield:2012aw}, \cite{Cordova:2015vwa} infer that for $SU(N)$ gauge group for any finite $N$ on $S^6$ we have the conformal anomaly
\bea
\int_{S^6} \A = \frac{2}{105} \cdot \(4N^3 - \frac{9}{4}N - \frac{7}{4}\) \frac{35}{8}
\eea

Let us now consider dimensions of curvature invariants. If we assign the metric tensor the length dimension $[g_{\mu\nu}] = 2$, then we get
\bea
[R^{\lambda}{}_{\mu\nu\rho}] &=& 0\cr
[R_{\mu\nu}] &=& 0\cr
[R] &=& - 2
\eea
Any product of these quantities or covariant derivatives thereof, such that all indices are contracted in the end, is called a local curvature invariant. We see that any local curvature invariant $K$ has an even dimension. If we integrate such a local curvature invariant over an $n$-dimensional manifold as $\int d^n x \sqrt{g} K$, then this integrated curvature invariant will have a dimension that is even if $n$ is even, and odd if $n$ is odd. 

Let us now assume that $[D] = -2$ and write $D = r^{-2} \h D$ where $[\h D] = 0$. Then we have the expansion
\bea
\tr(e^{-\frac{t}{r^2} \h D}) &=& \frac{1}{(4\pi)^3} \(\frac{a_0 r^6}{t^3} + \frac{a_2 r^4}{t^2} + \frac{a_4 r^2}{t} + a_6 + \O(t)\)
\eea
and we see that the HMDS coefficients have the following dimensions, $[a_0]=-6$, $[a_2]=-4$, $[a_4]=-2$ and $[a_6]=0$. If the HMDS coefficients are given by integrated local curvature invariants on a smooth manifold without boundary, then they must all have even dimensions. 

If we perform dimensional reduction down to five dimensions, the heat kernel expansion will acquire the following structure,
\bea
K(t) &=& \frac{1}{(4\pi)^{5/2}} \(\frac{a_0 r^5}{t^{5/2}} + \frac{a_2 r^3}{t^{3/2}} + \frac{a_4 r}{t^{1/2}} + a_5 + \O(t^{1/2})\)
\eea
If again we run the same sort of argument as above, we see that the HMDS coefficients have the dimensions $[a_0]=-5$, $[a_2]=-3$, $[a_4]=-1$ and $[a_5]=0$. If the HMDS coefficients are given by integrated local curvature invariants on a smooth five-manifold without boundary, then they must all have odd dimensions. So as $a_5$ has dimension $0$, which is even, we should find $a_5 = 0$. 

It is known how the HMDS coefficients $a_0$, $a_2$, $a_4$ and $a_6$ can be computed from curvature invariants \cite{Seeley:1967ea}, \cite{Gilkey:1975iq}. We can also compute these coefficients directly if we know the spectrum. Our method is based on the Euler-Maclaurin integral formula. This formula is normally used as an approximation method where a discrete sum is approximated by an integral. Our key observtion is that this approximation formula gives exact results for these first few heat kernel coefficients. This is a very general result. We then apply this method to the abelian M5 brane on $S^6$ where we can work out both the spectrum and the curvature invariants explicitly. Our first result is as follows. On $S^6$ we find agreement for all heat kernel coefficients $a_6$, for all fieds and ghosts fields that appear in the quantized $(2,0)$ tensor multiplet, when computed both ways, except for the ghost vector field where we need to add $1$ to the HMDS coefficient in order to match with the result that we get from the spectrum. This $1$ is later traced to an overcounted zero mode and is removed by hand. When we reduce along a singular fiber of $S^6$ down to 5d, we find that $a_5 = 0$ for all fields, except for the ghost vector field where we get $a_5 = 1$ that we later trace back to an overcounted zero mode that we remove by hand so as to get $a_5 = 0$ for all fields including the vector ghost. Not only $a_6$ in 6d has a physical interpretation but also the other heat kernel coefficients as well \cite{Vassilevich:2003xt}. For instance they contain the information about the short distance behavior of the propagotors. Our method gives exact results not only for $a_6$ but also for $a_0,...,a_5$. In section \ref{EML} we describe how we use the Euler-Maclaurin formula to compute $a_0,...,a_6$ exactly if we know the spectrum. In section \ref{6d} we apply this method to compute the heat kernel for the 6d tensor multiplet. In section \ref{5d} we perform dimensional reduction to 5d along a circle fiber that becomes singular at the north and south poles of $S^6$. In section \ref{discuss} we resolve the mismatch by removing any overcounted zero modes. There are three appendices. In appendix \ref{SO} we obtain the representations of $SO(7)$, Casimir invariants and dimensions corresponding to the various spherical harmonics on $S^6$, along with branching rules as we reduce along the singular fiber down to 5d. In appendix \ref{Tseytlin} we reproduce results in \cite{Bastianelli:2000hi} by applying the general formula in \cite{Gilkey:1975iq} to $S^6$. In appendix \ref{SUSY} we briefly discuss the partition function. We show that there is a huge supersymmetric cancelation of the modes.

This is a revised version where the mismatch that appeared in the first version has been resolved. I thank Arkady Tseytlin for pointing out relevant references where it was shown that this mismatch was due to a overcounting of zero modes in the heat kernel as we change variables.

\section{The heat kernel expansion}\label{EML}
Let us assume that $D$ is a differential operator on a six-manifold with eigenvalues $\lambda_n$ and degeneracies $d_n$. Let us further assume that we want to regularize the following, possibly divergent, partition function
\ben
Z_D = \frac{1}{\(\det D\)^{1/2}} = \prod_{n=0}^{\infty} \lambda_n^{-d_n/2}\label{Z}
\een
There are two different ways we may regularize. One way is to introduce the Minakshisundaram-Pleijel (MP) zeta function 
\ben
\zeta_D(s) &=& \sum_{n=0}^{\infty} d_n \lambda_n^{-s}\label{MPzeta}
\een
The other way is to introduce the heat kernel
\ben
K_D(t) &=& \sum_{n=0}^{\infty} d_n e^{-t \lambda_n}\label{Heat}
\een
We may define the partition function as
\bea
Z_D &=& e^{\frac{1}{2} \zeta'(0)}
\eea
Let us now assume that the eigenvalues take the form 
\bea
\lambda_n &=& \frac{\t{\lambda}_n}{r^2}
\eea
where $\t{\lambda}_n$ are dimensionless and $r$ is a length scale characterizing the six-manifold. The zeta function is
\bea
\zeta(s) &=& r^{2s} \t{\zeta}(s)\cr
\t{\zeta}(s) &=& \sum_{n=1}^{\infty} d_n \t{\lambda}_n^{-s}
\eea
and then
\bea
\zeta(0) &=& \t{\zeta}(0)\cr
\zeta'(0) &=& \t{\zeta}(0) \ln(r) + \t{\zeta}'(0)
\eea
and the partition function becomes 
\bea
Z_D &=& e^{\frac{1}{2} \t{\zeta}'(0)} r^{\frac{1}{2} \t{\zeta}(0)}
\eea
In order to see how the partition function scales with $r$, that is, the conformal anomaly, we only need to compute $\t{\zeta}(0)$, and not its derivative, which will be a slightly more complicated computation.

The two quantities are related by a Mellin transform as
\bea
\zeta_D(s) &=& \frac{1}{\Gamma(s)} \int_0^{\infty} dt t^{s-1} K_D(t)\cr
K_D(t) &=& \frac{1}{2\pi i} \oint_C ds t^{-s} \zeta_D(s) \Gamma(s)
\eea
where the contour encircles all the poles. If we only know the value of $\zeta_D(s)$ at $s = 0$, then we can only evaluate the inverse Mellin transform at the pole of $\Gamma(s)$ at $s=0$. Thus we can extract the following term in the heat kernel,
\bea
K_D(t) &=& \frac{1}{2\pi i} \oint_{s=0} \frac{ds}{s} \zeta(s) t^{-s} + ... = \zeta_D(0) + ...
\eea
What the above computation shows, is that 
\bea
a_6 &=& (4\pi)^3 \zeta_D(0)
\eea

The Mellin transform is defined as
\bea
\M K_D(s) &=& \int_0^{\infty} dt t^{s-1} K_D(t)
\eea
We have the relation 
\bea
\M K_D(s) &=& \zeta_D(s) \Gamma(s)
\eea
where 
\bea
\Gamma(s) &=& \int_0^{\infty} dx x^{s-1} e^{-x}
\eea
is the gamma function. The gamma function has simple poles at $s = - n$ for $n = 0,1,2,...$ with residues
\bea
\frac{1}{2\pi i} \oint_{s=- n} ds \Gamma(s) &=& \frac{(-1)^n}{n!}
\eea
The inverse Mellin transform is
\bea
K_D(t) &=& \frac{1}{2\pi i} \oint_C ds t^{-s} \M K_D(s)
\eea
where the contour encircles all the poles of $\M K_D(s)$. Let us check this formula by an example. Let us take $K_D(t) = e^{-t}$. The Mellin transform is $\M K_D(s) = \Gamma(s)$ and we get
\bea
K_D(t) = \frac{1}{2\pi i} \oint ds t^{-s} \Gamma(s) = \sum_{n=0}^{\infty} t^{n} \frac{(-1)^n}{n!} = e^{-t}
\eea
We have 
\bea
K_D(t) &=& \frac{1}{2\pi i} \oint_C ds t^{-s} \zeta(s) \Gamma(s)\cr
&=& \sum_{n=1}^{\infty} d_n \frac{1}{2\pi i} \oint ds t^{-s} \frac{1}{\lambda_n^s} \Gamma(s)
\eea
Now once having taking out the sum from the integral, the only poles are those of the gamma function. The integrals can be computed and we get the result
\bea
K_D(t) &=& \sum_{n=1}^{\infty} d_n \sum_{k=0}^{\infty} \(t \lambda_n\)^k  \frac{(-1)^k}{k!}\cr
&=& \sum_{n=1}^{\infty} d_n e^{-t \lambda_n}
\eea
This is a nice consistency check.

Let us now return to our general expression for the heat kernel, but instead of using the poles of the gamma function, let us this time use the poles from the MP-zeta function. Let us now suppose that we have the following pole structure for the MP-zeta function,
\bea
\zeta_D(s) &=& \frac{a_0}{s-3} + \frac{a_1}{s-2} + \frac{a_2}{s-1} + \zeta_{reg}(s) 
\eea
In particular, there is no pole at $s=0$, and near $s=0$ we have the expansion 
\bea
\zeta_D(s) &=& \zeta_D(0) + \O(s)
\eea
Then the heat kernel becomes
\bea
K_D(t) &=& a_0 \frac{1}{t^3} \Gamma(3) + a_1 \frac{1}{t^2} \Gamma(2) + a_2 \frac{1}{t} \Gamma(1) + \zeta_D(0) + \O(t)
\eea
Let us be more specific and let us assume that the eigenvalues and the degeneracies take the following form
\bea
\lambda_n &=& n^2 + 2 a n + b\cr
d_n &=& d_p n^p + d_{p-1} n^{p-1} + ...+ d_0
\eea
Then, by following closely the approach in \cite{Nash:1992sf}\footnote{By some more effort, this approach can also be used to compute the derivative $\zeta'(0)$.}, we may expand 
\bea
\frac{1}{\lambda_n^s} &=& \frac{1}{\(n^2 + 2 a n + b\)^s}\cr
&=& \frac{1}{n^{2s}} \frac{1}{\(1 + \frac{2a}{n} + \frac{b}{n^2}\)^s}\cr
&=& \frac{1}{n^{2s}} \(a_0 + \frac{a_{-1}}{n} + \frac{a_{-2}}{n^2} + \frac{a_{-3}}{n^3} + ...\)
\eea
and then we get the expansion
\bea
\zeta_D(s) &=& \sum_{n=1}^{\infty} \(d_p n^p + d_{p-1} n^{p-1} + ...+ d_0\) \cr
&& \frac{1}{n^{2s}} \(a_0 + \frac{a_{-1}}{n} + \frac{a_{-2}}{n^2} + \frac{a_{-3}}{n^3} + ...\)
\eea
We may write this as
\bea
\zeta_D(s) &=& \sum_{n=1}^{\infty} \(c_p n^{p-2s} + c_{p-1} n^{p-1-2s} + ... c_0 n^{-2s}+ c_{-1} n^{-1-2s} + ...\)
\eea
The coefficients are
\bea
c_p &=& d_p a_0\cr
c_{p-1} &=&  d_p a_{-1} + d_{p-1} a_0\cr
&...&\cr
c_0 &=& d_p a_{-p} + ... + d_0 a_0\cr
c_{-1} &=& d_p a_{-p-1} + ... + d_0 a_{-1}\cr
&...&
\eea
Now we can perform the sum over $n$ that gives
\bea
\zeta_D(s) &=& c_p \zeta(2s-p) + c_{p-1} \zeta(2s-p+1) + ... + c_0 \zeta(2s) + c_{-1} \zeta(2s+1) + ...
\eea
Finally we can evaluate this at $s=0$. 
\bea
\zeta_D(0) &=& c_p \zeta(-p) + c_{p-1} \zeta(1-p) + ... + c_0 \zeta(0) + \lim_{s\rightarrow 0} \(\frac{c_{-1}}{2s}\) 
\eea
Let us now assume that $p = 5$. Then we need the following coefficients
Here
\bea
a_0 &=& 1\cr
a_{-1} &=& \(- 2 a\) s + \O\(s^2\)\cr
a_{-2} &=& \(2 a^2 - b\) s + \O\(s^2\)\cr
a_{-3} &=& \(- \frac{8 a^3}{3} + 2 a b\)  s + \O\(s^2\)\cr
a_{-4} &=& \(4 a^4 - 4 a^2 b + \frac{b^2}{2}\) s + \O\(s^2\)\cr
a_{-5} &=& \(-\frac{32 a^5}{5} + 8 a^3 b - 2 a b^2\) s + \O\(s^2\)\cr
a_{-6} &=& \(\frac{32 a^6}{3} - 16 a^4 b + 6 a^2 b^2 - \frac{b^3}{3}\) s + \O\(s^2\)\cr
&...&
\eea
and then we get
\bea
\lim_{s\rightarrow 0} \frac{c_{-1}}{s} &=& d_5 \(\frac{32 a^6}{3} - 16 a^4 b + 6 a^2 b^2 - \frac{b^3}{3}\)\cr
&& + d_4 \(-\frac{32 a^5}{5} + 8 a^3 b - 2 a b^2\)\cr
&& + d_3 \(4 a^4 - 4 a^2 b + \frac{b^2}{2}\)\cr
&& + d_2 \(- \frac{8 a^3}{3} + 2 a b\)\cr
&& + d_1 \(2 a^2 - b\)\cr
&& + d_0 \(- 2 a\)
\eea
Also, since $a_{-n} = 0$ at $s=0$ for $n=1,2,...$, we see that at $s=0$ we have
\bea
c_p \zeta(-p) + c_{p-1} \zeta(1-p) + ... + c_0 \zeta(0) &=& d_p \zeta(-p) + d_{p-1} \zeta(1-p) + ... + d_0 \zeta(0)
\eea
We can now present the final result in a closed formula,
\ben
\zeta_{D}(0) &=& d_5 \(\zeta(-5) + \frac{16 a^6}{3} - 8 a^4 b + 3 a^2 b^2 - \frac{b^3}{6}\)\cr
&& + d_4 \(\zeta(-4) -\frac{16 a^5}{5} + 4 a^3 b - a b^2\)\cr
&& + d_3 \(\zeta(-3) + 2 a^4 - 2 a^2 b + \frac{b^2}{4}\)\cr
&& + d_2 \(\zeta(-2) - \frac{4 a^3}{3} + a b\)\cr
&& + d_1 \(\zeta(-1) + a^2 - \frac{b}{2}\)\cr
&& + d_0 \(\zeta(0) - a\)\label{zetaatnegative}
\een

If the spectrum is discrete, the trace of the heat kernel is an discrete sum. It may be hard to compute the sum exactly. However, a sum can be approximated by an integral using the Euler-Maclaurin formula \cite{M}. If we have a function $f(x)$ whose value at infinity and whose every derivative at infinity vanishes, then the Euler-Maclaurin formula can be reduced to 
\bea
\sum_{n=0}^{\infty} f(n) &=& \int_0^{\infty} dx f(x) + \frac{1}{2} f(0) - \sum_{k=1}^{\infty} \frac{B_{2k}}{(2k)!} f^{(2k-1)}(0)
\eea
where $B_{2k}$ are the Bernoulli numbers. It may seem that not much has been gained as the infinite sum on the right-hand side may seem to be even more difficult than the sum that we started with. Indeed, this is not the way that the Euler-Maclaurin formula is usually presented. Instead the sum on the right-hand side is usually truncated at some finite value and an error term is added. Here we are interested in exact results, and therefore it seems more appropriate for us to write the infinite sum without adding an error term. Here the derivatives at zero sit in a Taylor expansion of $f(x)$ around zero as
\bea
f(x) = \sum_{k=0}^{\infty} \frac{1}{k!} f^{(k)}(0) x^k = \sum_{k=0}^{\infty} d_k x^k
\eea
where we define 
\bea
d_k &=& \frac{1}{k!} f^{(k)}(0)
\eea
Then we get
\bea
\sum_{n=0}^{\infty} f(n) &=& \int_0^{\infty} dx f(x) + \frac{1}{2} f(0) - \sum_{k=1}^{\infty} \frac{B_{2k}}{2k} d_{2k-1}
\eea

Let us now apply this to the specific case when $D$ has eigenvalues on the form 
\bea
\lambda_n &=& n^2 + 2 a n + b
\eea
for some coefficients $a$ and $b$. Let us assume these eigenvalues come with the degeneracies
\bea
d_n &=& \sum_{k=0}^p d_k n^k
\eea
for some polynomial of degree $p$. The trace of the heat kernel is given by the sum
\bea
K(t) &=& \sum_{n=0}^{\infty} d_n e^{-t\(n^2 + 2 a n + b\)}
\eea
We then define the associated Euler-Maclaurin integral as
\bea
I(t) &=& \int_0^{\infty} dx d(x) e^{-t\(x^2 + 2 a x + b\)}
\eea
where
\bea
d(x) &=& \sum_{k=0}^p d_k x^k
\eea
The key point is that the full integrand when evaluated at $t=0$ reduces to $d(x)$, which is a polynomial of finite degree. It is this simple observation that explains how the Euler-Maclaurin formula can give us exact results for the first few coefficients in the small-$t$ expansion. By applying the Euler-Maclaurin formula, we find that
\bea
K(t) &=& I(t) + \frac{1}{2} d_0 - \sum_{k=1}^p \frac{B_k}{k} d_{k-1} + \O(t)
\eea
For $p=5$ we get
\bea
K(t) &=& I(t) + \frac{1}{2} d_0 - \frac{1}{12} d_1 + \frac{1}{120} d_3 - \frac{1}{252} d_5  
\eea
where the integral can be computed. Its series expansion reads
\bea
I(t) &=& \sum_{k=0}^{\infty} \frac{a_k}{t^{(6-k)/2}}
\eea
where
\bea
a_0 &=& d_5\cr
a_2 &=& \(6a^2 - b\) d_5 - 2 a d_4 + \frac{1}{2} d_3\cr
a_4 &=& \(16a^4-12a^b+b^2\)d_5 + \(4ab-8a^3\)d_4 + \(4a^2-b\)d_3 - 2a d_2 + d_1\cr
a_6 &=& \(\frac{16}{3}a^6 - 8a^4 b + 3 a^2 b^2 - \frac{1}{6} b^3\) d_5 + \(-\frac{16}{5}a^5 + 4 a^3 b - a b^2\) d_4\cr
&& + \(2 a^4 - 2 a^2 b + \frac{1}{4} b^2\) d_3 + \(-\frac{4}{3} a^3 + ab\) d_2\cr
&& + \(a^2 - \frac{1}{2} b\) d_1 - a d_0
\eea
We also find that 
\bea
a_1 &=& \frac{3\sqrt{\pi}}{8} \(d_4 - 5 a d_5\)\cr
a_3 &=& \frac{\sqrt{\pi}}{8} \(\(15ab-35a^3\)d_5+\(15a^2-3b\)d_4-6ad_3+2d_2\)\cr
a_5 &=& \frac{\sqrt{\pi}}{16} \Bigg(\(-63a^5+70a^3b-15ab^2\)d_5+\(35a^4-30a^2b+3b^2\)d_4\cr
&&+\(-20a^3+12ab\)d_3+\(12a^2-4b\) d_2 - 8a d_1 + 8 d_0\Bigg)
\eea
The heat kernel coefficients $a_k$ are integrals of local geometric invariants. No such invariant exists for a smooth manifold without boundary in odd dimensions. Therefore we must have $a_k = 0$ for odd $k$'s. This puts constraints the spectrum, the degeneracy must be correlated with the eigenvalues so that $a_k = 0$ for odd $k$'s. For all the cases that we will encounter, we find that indeed $a_k = 0$ for odd $k$'s.

By identifying this result with the one that we got in (\ref{zetaatnegative}), we find the identity
\bea
\sum_{k=0}^p \zeta(-k) d_k &=& - \frac{1}{2} d_0 - \frac{1}{12} d_1 + \frac{1}{120} d_3 - \frac{1}{252} d_5
\eea
which gives us the values of the Riemann zeta function at negative integers \cite{DG}, \cite{AGB}, \cite{KB}
\bea
\zeta(-k) &=& (-1)^k \frac{B_{k+1}}{k+1}
\eea

\section{The heat kernel for the tensor multiplet}\label{6d}
The 6d $(2,0)$ tensor multiplet consists of five scalars $\phi^A$, a two-form $B_{MN}$ with selfdual field strength $H_{MNP} = \partial_M B_{NP} + \partial_N B_{PM} + \partial_P B_{MN}$, and four $SO(5)$-Majorana-Weyl fermions $\psi$. The $SO(5)$-Majorana condition means 11d Majorana spinor reduced to 6d, where a 6d Weyl projection is imposed. The superconformal Lagrangian on a Lorentzian six-manifold is given by
\bea
\L &=& \frac{1}{2\pi} \(-\frac{1}{24} H_{MNP}^2 - \frac{1}{2} (D_M\phi^A)^2 - \frac{R}{10} (\phi^A)^2 + \frac{i}{2} \bar\Psi\Gamma^M D_M\Psi\)
\eea
where $R$ is the Ricci scalar. In this Lagrangian, $H_{MNP}$ is non-selfdual. For the quantization of the two-form gauge field, we need to supplement this Lagrangian with terms coming from two anticommuting vector-ghosts and three commuting massless scalar ghosts. The partition 
function for a non-selfdual two-form, including the ghost contributions, is given by
\bea
Z_B &=& \frac{\det\triangle_1}{\det^{\frac{1}{2}}\triangle_2 \det^{\frac{3}{2}}\triangle_0}
\eea
By Hodge decomposition, any nonzero mode is either exact or coexact. We may then write  the partition function as \cite{Bak:2016vpi}
\bea
Z_B &=& \frac{\det^{\frac{1}{2}}\triangle_1^{coex}}{\det^{\frac{1}{2}}\triangle_2^{coex} \det^{\frac{1}{2}}\triangle_0^{coex}}
\eea
where the subscript $coex$ means that we restrict to the space of coexact forms. On $S^6$ it is these coexact form that correspond to states in irreducible representations of $SO(7)$. 

We will now apply the Euler-Maclaurin formula to obtain the heat kernel expansions for each field in the tensor multiplet. The sum that we compute is on the general form
\bea
K(t) &=& \sum_{n=0}^{\infty} d_n e^{-t \lambda_n}
\eea
To compute the small-t expansion of this sum, we use the spectrum ($\lambda_n$ and $d_n$) of the corresponding differential operator acting on that field. We work out the spectrum on $S^6$ using representation theory of its isometry group $SO(7)$ in appendix \ref{SO}.

\subsection{Conformally coupled scalar field}
The Ricci scalar is $R = 30$ on $S^6$ of unit radius. The conformal Laplacian $\triangle_{conformal} = \triangle + 6$ has the spectrum 
We have
\bea
\lambda_n &=& n^2 + 5 n + 6\cr
d_n &=& \frac{1}{60} n^5 + \frac{5}{24} n^4 + n^3 + \frac{55}{24} n^2 + \frac{149}{60} n + 1
\eea
The Euler-Maclaurin integral is 
\bea
I(t) &=& e^{-6t} \int_0^{\infty} s(x) e^{-t(x^2+5x)}
\eea
It has the series expansion
\bea
I(t) &=& \frac{1}{60} \(\frac{1}{t^3} - \frac{1}{t^2} - 18 + \O(t)\)
\eea
The Bernoulli part is
\bea
\frac{1}{2} d_0 - \frac{1}{12} d_1 + \frac{1}{120} d_3 - \frac{1}{252} d_5 &=& \frac{1139}{3780}
\eea
In total therefore
\bea
a_6 = \frac{1139}{3780} - \frac{18}{60} = \frac{1}{756}
\eea

\subsection{Massless scalar ghost}
We have
\bea
\lambda_n &=& n^2 + 5 n\cr
d_n &=& \frac{1}{60} n^5 + \frac{5}{24} n^4 + n^3 + \frac{55}{24} n^2 + \frac{149}{60} n + 1
\eea
The Euler-Maclaurin integral is 
\bea
I(t) &=& \frac{1}{60} \(\frac{1}{t^3} + \frac{5}{t^2} + \frac{12}{t}\)
\eea
The Bernoulli contribution is 
\bea
\frac{1}{2} d_0 - \frac{1}{12} d_1 + \frac{1}{120} d_3 - \frac{1}{252} d_5 &=&  \frac{1139}{3780}
\eea
In total therefore
\bea
a_6 = \frac{1139}{3780} 
\eea

\subsection{Vector ghost}
We have
\bea
\lambda_n &=& n^2 + 5n + 4\cr
d_n &=& \frac{1}{12} n^5 + \frac{25}{24} n^4 + \frac{14}{3} n^3 + \frac{215}{24} n^2 + \frac{25}{4} n
\eea
The Euler-Maclaurin integral is
\bea
I(t) &=& e^{-4t} \frac{1+3t}{12 t^3}
\eea
Its series expansion is 
\bea
I(t) &=& \frac{1}{12 t^3} - \frac{1}{12 t^2} - \frac{1}{3t} + \frac{10}{9} + \O(t)
\eea
The Bernoulli part is 
\bea
\frac{1}{2} d_0 - \frac{1}{12} d_1 + \frac{1}{120} d_3 - \frac{1}{252} d_5 &=& - \frac{1823}{3780}
\eea
In total therefore
\bea
a_6 = \frac{10}{9} - \frac{1823}{3780} = \frac{2377}{3780}
\eea
From $a_0$ we read off that the vector field here has 5 components (since $1/12 = 5/60$). This can be understood from that co-exact one-forms constitute only one part of the vector harmonics. The other part is consisting of one derivative acting on scalar harmonics. That amounts to in total $5+1=6$ vector components.

\subsection{Two-form gauge field}
We have
\bea
\lambda_n &=& n^2 + 5 n + 6\cr
d_n &=& \frac{n^5}{6} + \frac{25 n^4}{12} + 9 n^3 + \frac{185 n^2}{12} + \frac{25 n}{3}
\eea
The Euler-Maclaurin integral is 
\bea
I(t) &=& e^{-6t} \frac{1+2t}{6t^3}
\eea
whose series expansion is 
\bea
I(t) &=& \frac{1}{6t^3} - \frac{2}{3t^2} + \frac{1}{t} + \O(t)
\eea
The Bernoulli contribution is 
\bea
\frac{1}{2} d_0 - \frac{1}{12} d_1 + \frac{1}{120} d_3 - \frac{1}{252} d_5 &=& - \frac{586}{945}
\eea
In total therefore
\bea
a_6 &=& - \frac{586}{945}
\eea
From $a_0$ we read off that the two-form has $10 = 5 \cdot 4/2$ components since $1/6 = 10/60$. 

\subsection{Fermion} 
The fermionic harmonics correspond to the representation $(n,0,1)$ of $SO(7)$. This representation has the Casimir invariant and the dimension 
\bea
\lambda^{Casimir}_n &=& (n+3)^2 - \frac{15}{4}\cr
d_n &=& \frac{n^5}{15} + \frac{2n^4}{3} + \frac{7n^3}{3} + \frac{10 n^2}{3} + \frac{8 n}{5}
\eea
According to \cite{Trautman:1995fr}, the eigenvalues of the Dirac operator on $S^6$ are $\pm(n+3)$ where each sign comes with the degeneracy $f_n = \frac{1}{15}(n+1)(n+2)(n+3)(n+4)(n+5)$ for $n=0,1,2,...$. We see that $f_n = d_n$, while the eigenvalues are related to the Casimir invariant through the Lichnerowicz formula
\bea
\(i \Gamma^{\mu} D_{\mu}\)^2 &=& - D^{\mu} D_{\mu} - \frac{R}{8} 
\eea
where for $S^6$ we have $R/8 = 15/4$. The partition function is
\bea
Z_F = \prod_{n=0}^{\infty} \(n+3\)^{2 f_n} = \prod_{n=0}^{\infty} \(n^2 + 6 n + 9\)^{f_n}
\eea
The associated heat kernel is 
\bea
- 2 \sum_{n=0}^{\infty} f_n e^{-t \(n^2 + 6 n + 9\)}
\eea
The overall factor of $-2$ comes from the way that we map (\ref{Z}) to (\ref{Heat}). We have the Euler-Maclaurin integral
\bea
I(t) &=& - e^{-9t} \frac{2+13 t+40t^2}{15 t^3}
\eea
whose series expansion is 
\bea
I(t) &=& - \(\frac{2}{15 t^3} - \frac{1}{3 t^2} + \frac{4}{15 t} - \frac{51}{10} + \O(t)\)
\eea
The Bernoulli contribution is 
\bea
-2\(\frac{1}{2} d_0 - \frac{1}{12} d_1 + \frac{1}{120} d_3 - \frac{1}{252} d_5\) &=&  - \frac{19087}{3780}
\eea
In total therefore
\bea
a^F_6 = \frac{51}{10} - \frac{19087}{3780} = \frac{191}{3780} 
\eea
From $2/15 = 8/60$ we get $8$ spinor components. In the 6d theory we have 4 real Weyl spinors. This has the same number of components as 2 real 8-component Dirac spinors or one complex 8-component Dirac spinor.

\subsection{The $(2,0)$ tensor multiplet}
We summarize the above results. For the conformal scalar ($S$), massless scalar ghost ($S_0$), vector ghost ($V$), two-form ($T$) and fermion ($F$) respectively, we have
\bea
a^S_6&=& \N \frac{5}{72}\cr
a^{S_0}_6 &=& \N \frac{1139}{72}\cr
a^V_6 &=& \N \frac{2377}{72}\cr
a^T_6 &=& - \N \frac{586}{18}\cr
a^F_6 &=& \N \frac{191}{72}
\eea
where we have taken out a common factor
\bea
\N &=& \frac{2}{105}
\eea
The heat kernel coefficient associated to the two-form is 
\bea
a^B_6 &=& a^T_6 - a^V_6 + a^{S_0}_6
\eea
because spherical harmonics correspond to co-exact forms \cite{Bak:2016vpi}. We get
\bea
a^B_6 &=& \N \frac{221}{4} - 2
\eea
If we use the total number of ghosts in the ghost tower, then we have the relation
\bea
a^B_6 &=& b_6^{tot,T} - 2 b_6^{tot,V} + 3 b_6^{tot,S_0}
\eea
corresponding to $2$ vector ghosts and $3$ massless scalar ghosts. The coefficients are related with those above as
\bea
a^{tot,T}_6 &=& a^{T}_6 + a^{V}_6\cr
a^{tot,V}_6 &=& a^{V}_6 + a^{S_0}_6\cr
a^{tot,S_0}_6 &=& a^{S_0}_6
\eea
We get
\bea
a_6^{tot,T} &=& \N \frac{11}{24}\cr
a_6^{tot,V} &=& \N \frac{293}{6}\cr
a_6^{tot,S_0} &=& \N \frac{1139}{72}
\eea
In the tensor multiplet there are $5$ conformally coupled scalar fields, one selfdual two form and $4$ Majorana-Weyl fermions. The heat kernel coefficient for the tensor multiplet is therefore
\bea
a^{M5}_6 = 5a^S_6 + \frac{1}{2} a^B_6 + a^F_6 = - \frac{5}{12}
\eea

Let us now compare our result with the coefficients\footnote{We ignore the overall factor $-\frac{1}{(4\pi)^3 7!}$ in their expressions.} of the Euler density that were obtained in \cite{Bastianelli:2000hi},
\bea
a^S &=& \frac{5}{72}\cr
a^B &=& \frac{221}{4}\cr
a^F &=& \frac{191}{72}
\eea
For the whole tensor multiplet then 
\bea
a^{M5} = 5 S + \frac{1}{2} B + F = - \frac{245}{8}
\eea
We see that we have a perfect match\footnote{We can not really say that much because we could always fix say $a^S$ to whatever number we like by  changing the overall coefficient. However, we think our match is much stronger than that. One reason to believe so, is because $191$ is a large prime integer number.} for $a^S$ and $a^F$. Then we have a mismatch only for $a^B$. We would now like to track this mismatch a bit further. In \cite{Bastianelli:2000hi}, the explicit expressions for the ghosts associated to the two-form were not written down in the final form where the coefficient of $E_6$ could have been read off. But we can easily extract the value of that coefficient from the expressions presented in \cite{Bastianelli:2000hi} by evaluating the curvature invariants\footnote{The relations between curvature invariants in \cite{Bastianelli:2000hi} and our curvature invariants (\ref{curv}) are $A_{10} = -L_1, A_{11} = -L_2, A_{12} = -L_3$ and $A_{13} = -K_1,A_{14} =K_2,
A_{15} = -K_3, A_{16} = K_4, A_{17} = K_5$. The various minus signs arise from our convention that $R_{ij} = R_{kijk} = - R_{kikj}$} on $S^6$. By using the expression of $\A_B$ written in terms of curvature invariants\footnote{Again we ignore the prefactor $-\frac{1}{(4\pi)^3 7!}$}, we get
\bea
\A_B = \frac{442}{7} = \frac{8}{7} \cdot \frac{221}{4}
\eea
Hence there appears (no surprise) the factor of $\frac{8}{7} = 60 \N$ that shall be common to all expressions in \cite{Bastianelli:2000hi} where curvature invariants appear. We further compute those expressions in \cite{Bastianelli:2000hi} for the individual contributions to the two-form on $S^6$ with the following results
\bea
a^{T} &=& \frac{8}{7}\cdot \frac{11}{24}\cr
a^{V} &=& \frac{8}{7}\cdot \(-\frac{11}{3}\)\cr
a^{S_0} &=& \frac{8}{7}\cdot \frac{1139}{72}
\eea
Comparing these numbers with ours, we see that there is mismatch only for $a^{tot,V}_6$. Re-instating the normalization factor $\N$, this mismatch becomes 
\ben
a^{tot,V}_6 - a^V = \N \cdot \(\frac{293}{6} + \frac{11}{3}\) = 1\label{6dV}
\een

\section{Dimensional reduction to five dimensions}\label{5d}
We view $S^6$ as an $S^5$ fibered over an interval such that $S^5$ shrinks to zero size at the end-points, and perform dimensional reduction along the Hopf fiber of $S^5$. Group theoretically this amounts to first re-arranging the $SO(7)$ harmonics in terms of $SO(6)$ harmonics of $S^5$, and subsequently reducing $SO(6) = SU(4) \rightarrow SU(3) \times U(1)_H$ where $U(1)_H$ is the isometry group of the Hopf circle. Dimensional reduction amounts to keeping only the modes that are neutral under $U(1)_H$. 

\subsection{Massless scalar ghost}
We will now pick the states with zero $U(1)_H$ charge from the heat kernel of scalar harmonics $(n,0,0)$. Using the branching rules in the appendix \ref{SO}, we see that we shall keep the following representations of $SU(3)$,
\bea
R_{2m} &=& \bigoplus_{\l=0}^m (\l,\l)\cr
R_{2m+1} &=& \bigoplus_{\l=0}^m (\l,\l)
\eea
We have the dimension $\dim(\l,\l) = (\l+1)^3$. Then the heat kernel becomes 
\bea
K_{S_0}(t) &=& \sum_{m=0}^{\infty} \(e^{- t(4m^2+8m)}+e^{- t(4m^2+12m+5)}\) \sum_{\l=0}^m (\l+1)^3  
\eea
The sum is 
\bea
\sum_{\l=0}^m (\l+1)^3 &=& \frac{1}{4} (m+1)^2 (m+2)^2
\eea
The corresponding Euler-Maclaurin integral has the series expansion
\bea
&&\int_0^{\infty} dx \frac{1}{4}(x+1)^2(x+2)^2 \(e^{- t(4x^2+10x)} + e^{-t(4x^2+14 x+6)}\) \cr
&=& \frac{3\sqrt{\pi}}{512}\(\frac{1}{t^{5/2}} + \frac{71}{12} \frac{1}{t^{3/2}} + \frac{1747}{96}\frac{1}{t^{1/2}}\) - \frac{21}{40} + \O(t^{1/2})
\eea
The Bernoulli contribution is obtained from expanding out the summand at $t=0$
\bea
\frac{1}{2} (m+1)^2 (m+2)^2 &=& \frac{m^4}{2} + 3m^3 + \frac{13 m^2}{2} +6 m + 2
\eea
Then 
\bea
\frac{1}{2} d_0 - \frac{1}{12} d_1 + \frac{1}{120} d_3 - \frac{1}{252} d_5 = \frac{1}{2}\cdot 2 - \frac{1}{12} \cdot 6 + \frac{1}{120} \cdot 3 = \frac{21}{40}
\eea
Thus we get
\bea
a_5^{S_0} = -\frac{21}{40} + \frac{21}{40} = 0
\eea

\subsection{Conformally coupled scalar}
Let us next consider the conformally coupled scalar. This gives the Euler-Maclaurin integral
\bea
I_S(t) &=& \frac{3\sqrt{\pi}}{512} \(\frac{1}{t^{5/2}} - \frac{1}{12t^{3/2}} + \frac{67}{96 t^{1/2}}\) - \frac{21}{40}\cr
&=& \frac{3\sqrt{\pi}}{512} \(\frac{1}{t^{5/2}} + \(\frac{71}{12} - 6\) \frac{1}{t^{3/2}} + \frac{67}{96 t^{1/2}}\) - \frac{21}{40}
\eea
while the Bernoulli part is the same as for the massless scalar, resulting in 
\bea
a_5^{S} &=& 0
\eea

The other heat kernel coefficients are of course interesting to study as well. From $a_0$ and $a_2$ we may deduce that
\bea
\Vol &=& \frac{3\sqrt{\pi}}{512}\cr
R &=& \frac{71}{2}
\eea
For $a_4$, we may understand the difference by applying a formula for $a_4$ where for a conformal scalar $E = - 6$,
\bea
\frac{1}{360} \(60 RE + 180 E^2\) &=& - \frac{35}{2}
\eea
We next notice that 
\bea
\frac{67}{96} - \frac{1747}{96} &=& - \frac{35}{2}
\eea
This explains the difference betweeen $a_4$ in for the above the two heat kernels.

\subsection{Vector ghost}
The representations with zero $U(1)_H$ charge can be extracted from the branching rules in the appendix \ref{SO}. They are
\bea
R_{n=2m} &=& \bigoplus_{\l=1}^m (\l,\l) \oplus \bigoplus_{\l=0}^{m-1} \((\l,\l)\oplus (\l+2,\l-1)\oplus (\l-1,\l+2)\oplus (\l+1,\l+1)\)\cr
R_{n=2m+1} &=& \bigoplus_{\l=1}^m (\l,\l) \oplus \bigoplus_{\l=0}^{m} \((\l,\l)\oplus (\l+2,\l-1)\oplus (\l-1,\l+2)\oplus (\l+1,\l+1)\)
\eea
with the corresponding dimensions
\bea
\dim R_{2m} &=& \frac{5}{4}m^4+\frac{11}{2}m^3+\frac{29}{4}m^2+3m\cr
\dim R_{2m+1} &=& \frac{5}{4}m^4+\frac{19}{2}m^3+\frac{101}{4}m^2+27m+9
\eea
The heat kernel becomes
\bea
K(t) &=& \sum_{m=0}^{\infty} \(d^0_{2m} e^{-t \lambda_{2m}} + d^0_{2m+1} e^{-t \lambda_{2m+1}}\)
\eea
The corresponding Euler-Maclaurin integral becomes
\bea
I(t) &=& \frac{3\sqrt{\pi}}{512} \(\frac{5}{t^{5/2}} - \frac{77}{12 t^{3/2}} - \frac{3521}{96 t^{1/2}}\) - \frac{9}{8} + \O(t^{1/2})
\eea
and the Bernouill part is extracted from the summand at $t=0$,
\bea
d_{2m} + d_{2m+1} &=& \frac{5}{4} m^4 + 15 m^3 + \frac{65}{2} m^2 + 30 m + 9
\eea
and becomes
\bea
\frac{1}{2} d_0 - \frac{1}{12} d_1 + \frac{1}{120} d_3 &=& \frac{17}{8}
\eea
Thus in total we get
\bea
a_5^V = \frac{17}{8} - \frac{9}{8} = 1
\eea

\subsection{Two-form gauge field}
The representations with zero $U(1)_H$ charge can be extracted from the branching rules in the appendix \ref{SO}. They are
\bea
R_{n=2m} &=& \bigoplus_{\l=2}^m (\l+1,\l-2)^2 \oplus \bigoplus_{\l=1}^m (\l,\l)^2 \oplus \bigoplus_{\l=0}^m (\l-1,\l+2)^2\cr
&&\oplus \bigoplus_{\l=0}^{m-1} \((\l,\l)\oplus (\l+2,\l-1)\oplus (\l+1,\l+1) \oplus (\l+1,\l)\)\cr
R_{n=2m+1} &=& \bigoplus_{\l=2}^m (\l+1,\l-2)^2 \oplus \bigoplus_{\l=1}^m (\l,\l)^2 \oplus \bigoplus_{\l=0}^m (\l-1,\l+2)^2\cr
&&\oplus \bigoplus_{\l=0}^{m} \((\l,\l)\oplus (\l+2,\l-1)\oplus (\l+1,\l+1) \oplus (\l+1,\l)\)
\eea
with the corresponding dimensions
\bea
\dim R_{2m} &=& \frac{9}{4} m^4 + 12 m^3 + \frac{81}{4} m^2 + \frac{21}{2} m\cr
\dim R_{2m+1} &=& \frac{9}{4} m^4 + 15 m^3 + \frac{135}{4} m^2 + 30 m + 9
\eea
The heat kernel becomes
\bea
K(t) &=& \sum_{m=0}^{\infty} \(d^0_{2m} e^{-t \lambda_{2m}} + d^0_{2m+1} e^{-t \lambda_{2m+1}}\)
\eea
We get
\bea
I(t) &=& \frac{27\sqrt{\pi}}{512} \(\frac{1}{t^{5/2}} - \frac{49}{12 t^{3/2}} + \frac{419}{96 t^{1/2}}\) - \frac{27}{20} + \O(t^{1/2})
\eea
and from the summand at $t=0$
\bea
\frac{9}{2} m^4 + 27 m^3 + 54 m^2 + \frac{81}{2} m + 9
\eea
we read off that 
\bea
\frac{1}{2} d_0 - \frac{1}{12} d_1 + \frac{1}{120} d_3 &=& \frac{27}{20}
\eea
Thus 
\bea
a_5^T = -\frac{27}{20} + \frac{27}{20} = 0
\eea

\subsection{Fermion}
We see that no states are neutral under $U(1)_H$ and we get
\bea
K(t) &=& 0
\eea
and so trivially we have 
\bea
a_5^F &=& 0
\eea

\subsection{The 5d SYM}
The total heat kernel coefficient is obtained by the same formula as in 6d,
\bea
a_5^{SYM} &=& 5 a_5^{S} + \frac{1}{2} a_5^B + a_5^F,\cr
a_5^B &=& a_5^T - a_5^V + a_5^{S_0}
\eea
where, from the above results, we have
\bea
a_5^S &=& 0\cr
a_5^{S_0} &=& 0\cr
a_5^{V} &=& 1\cr
a_5^T &=& 0\cr
a_5^F &=& 0
\eea
Therefore we get
\bea
a_5^{SYM} &=& \frac{1}{2}
\eea
This is not an integer and seems to be not directly related to the mismatch of $1$ that we got in 6d. But let us notice that for the total heat kernels
\bea
a_5^{tot,T} &=& a_5^T + a_5^V\cr
a_5^{tot,V} &=& a_5^V + a_5^{S_0}\cr
a_5^{tot,S_0} &=& a_5^{S_0}
\eea
we have
\bea
a_5^{tot,T} &=& 1\cr
a_5^{tot,V} &=& 1\cr
a_5^{tot,S_0} &=& 0
\eea
Let us further note that a two-form in 6d, reduces to both a two-form and a one-form in 5d. So it seems that we always find this extra contribution of $1$ from the one-forms, wherever they appear.

\section{Resolving the mismatch}\label{discuss}
Let us now resolve the mismatch. 

There is a zero mode in the spectrum. The zero mode is for the scalar ghost where $\lambda_n = n^2 + 4 n = 0$ for $n = 0$ which comes with the degeneracy $d_{0} = 1$. Thus here is a zero mode that we need to further gauge fix. In the end that gauge fixing amounts to just removing the $n=0$ mode from the spectrum. It does not concern the vector ghost, and at first sight it does not seem to explain the mismatch. One may argue that we should then shift $a_6^{S_0}$ by one unit. If we do that, then we get one more mismatch. It does not cure the mismatch. It makes it worse. The heat kernel is well-defined with the zero mode included. In fact, zero modes play an important role in heat kernels when they are applied to index theorems \cite{Gilkey}. 

Nevertheless, the key to understanding the mismatch, is to realize that the heat kernel includes zero modes, and that new zero modes can arise as we do the Hodge decomposition of a $p$-form. Any such new zero modes that arises by the Hodge decomposition must be removed by hand since they were not there originally. This mechanism was first discovered in \cite{Christensen:1979iy}, \cite{Fradkin:1983mq} and it was applied to higher spin fields on $S^6$ in \cite{Tseytlin:2013fca}. I would like to thank Tseytlin for helping me resolve the mismatch puzzle by pointing out the relevant references. 

Let us now illustrate how this mechanism resolves the mismatch by considering the vector ghost. The vector field $v_i$ on $S^6$ decomposes into an exact plus a coexact piece as
\bea
v_i &=& v'_i + \partial_i v
\eea
for the non-zero modes. This decomposition amounts to the relation
\bea
\det{}' \triangle^{tot}_1 &=& \det{}' \triangle^{coex}_1 \det{}' \triangle_0^{coex}
\eea
where zero modes are taken out, as indicated by primes. These determinants may be written in terms of the MP zeta functions as
\bea
e^{-\zeta'_{\triangle^{tot}_1}(0)} &=& e^{-\zeta'_{\triangle_1^{coex}}(0)} e^{-\zeta'_{\triangle^{coex}_0}(0)}
\eea
where zero modes are not included. The relation for the heat kernels is different because for the heat kernels we include the zero modes. Therefore, for the heat kernels, the corresponding relation will read
\bea
K^{tot}_{\triangle_1}(t) &=& K_{\triangle_1^{coex}}(t) + \(K_{\triangle_0^{coex}}(t) - 1\)
\eea
where for the scalar $v$ we have to subtract the zero modes that otherwise would be overaccounted for, since it was not there originally, on the left-hand side. Namely, these zero modes do not survive as we take the derivative of $v$ to get the exact piece $v_i = \partial_i v + ...$. There is exactly one zero mode for a scalar on $S^6$. This amounts to a correction of our previous claim that $a_6^{tot,V} = a_6^V + a^{S_0}_6$. The corrected relation reads
\bea
a_6^{tot,V} = a^V_6 + \(a^{S_0}_6 - 1\) 
\eea
This in turn corrects (\ref{6dV}) to now read
\bea
a_6^{tot,V} - a^V &=& 0
\eea
and thus after the correction, we find agreement. Let us next consider the two-form. Again Hodge-decomposing the two-form in a coexact and exact pieace as
\bea
B_{ij} &=& B'_{ij} + \partial_{[i} B_{j]}
\eea
we find that we need to subtract any zero modes of the vector field $B_i$ on $S^6$ from the heat kernel for the two-form. Now there are no zero modes for the vector field on $S^6$, so therefore we got the agreement for the two-form without any correction.

As we dimensionally reduce, we again find the same pattern. By correcting for zero mode for the ghost vector, we find the result $a_5^V = 0$.

Our result can not be used to exclude the possibilty that we may need to add some extra local degrees of freedom at loci where the circle fiber shrinks to zero size, that is, at the north and south poles of $S^6$. Such extra local contributions will not depend on $r$ because the local geometry near the norh and south poles is flat $\mb{R}^5$ and no $r$-dependence can arise from there. A similar situation occurs when we put M5 brane on $S^4 \times \Sigma$ where $\Sigma$ is a Riemann manifold \cite{Cordova:2016cmu}. If we view $S^4$ as $S^3$ fibered over an interval such that $S^3$ shrinks to zero size at the end points, and reduce along the Hopf fiber of $S^3$, we find new degrees of freedom at the north and south poles corresponding to a D4 brane ending on a D6 brane. We may remove the north pole of $S^4$ by cutting along a small boundary-$S^3$ near the north pole (and similarly for the south pole). In M-theory we then have the boundary manifold $S^3 \times N_5$ where $N_5$ is the five-dimensional normal bundle of $S^4 \times \Sigma$ in eleven dimensions. We see that by cutting out the north pole, the M5 brane will end on an eight-manifold. Upon dimensional reduction along the Hopf fiber of $S^3$ down to Type IIA string theory, this eight-manifold becomes a seven-manifold that will correspond to a D6-brane as was shown in \cite{Cordova:2016cmu}. For our $S^6$ we may do an analogous thing. Cutting along a small $S^5$ near the north pole we get a boundary manifold $S^5 \times N_5$. This boundary should correspond to an M9 brane in M-theory. We may reduce along the Hopf fiber of $S^5$ down to Type IIA string theory, where we find D4 brane ending on some nine-manifold, which should be identified with some 8-brane in IIA string theory on which D4 can end.

\subsection*{Acknowledgments}
I would like to Arkady Tseytlin for helping me resolve the mismatch puzzle, and Dongsu Bak for discussions. This work was supported in part by NRF Grant 2017R1A2B4003095.

\appendix 
\section{Representations of $SO(7)$}\label{SO}
We follow the reference \cite{Cahn} to study representations of the Lie algebra $B_3$ of the isometry group $SO(7)$ of $S^6$. The Cartan matrix is defined as
\bea
A_{ij} &=& \frac{2\<\alpha_i,\alpha_j\>}{\<\alpha_j,\alpha_j\>}
\eea
and for $B_3$ this is given by 
\bea
A_{ij} &=& \(\begin{array}{ccc}
2 & -1 & 0\\
-1 & 2 & -2\\
0 & -1 & 2
\end{array}\)
\eea
The inverse is
\bea
A^{ij} &=& \frac{1}{2} \(\begin{array}{ccc}
2 & 2 & 2\\
2 & 4 & 4\\
1 & 2 & 3
\end{array}\)
\eea
Dynkin coefficients of a weight $\Lambda$ are defined as
\bea
\Lambda_i &=& \frac{2\<\Lambda,\alpha_i\>}{\<\alpha_i,\alpha_i\>}
\eea
The Dynkin coefficients of the simple root $\alpha_i$ become
\bea
(\alpha_i)_j = \frac{2\<\alpha_i,\alpha_j\>}{\<\alpha_j,\alpha_j\>} = A_{ij}
\eea
that is, the $i$-th row in the Cartan matrix is the simple root $\alpha_i$ in Dynkin labels. Thus the simple roots have the Dynkin labels 
\bea
\alpha_1 &=& (2,-1,0)\cr
\alpha_2 &=& (-1,2,-2)\cr
\alpha_3 &=& (0,-1,2)
\eea
The $9$ positive roots are $\{\alpha_1,\alpha_2,\alpha_3,\alpha_1 + \alpha_2,\alpha_2 + \alpha_3,\alpha_2 + 2 \alpha_3,\alpha_1 + \alpha_2 + \alpha_3,
\alpha_1 + \alpha_2 + 2 \alpha_3,\alpha_1 + 2 \alpha_2 + 2 \alpha_3\}$. The sum of them is 
\bea
2\delta &=& 5\alpha_1+8\alpha_2+9\alpha_3
\eea 
The Casimir invariant associated with the representation with the highest weight $\Lambda$ is given by (see equation (XI.23) in \cite{Cahn}) 
\bea
C(\Lambda) &=& \<\Lambda,\Lambda\> + \<\Lambda,2\delta\>
\eea
The dimension of this representation is given by Weyl's dimension formula
\bea
dim(\Lambda) &=& \prod_{\alpha>0} \frac{\<\alpha,\Lambda+\delta\>}{\<\alpha,\delta\>}
\eea
Explicitly, we get
\bea
\dim(\Lambda) = &\frac{1}{720}&(\Lambda_1+1)(\Lambda_2+1)(\Lambda_3+1)\cr
&&(\Lambda_1+\Lambda_2+2)(\Lambda_2+\Lambda_3+2)(2\Lambda_2+\Lambda_3+3)\cr
&&(\Lambda_1+2\Lambda_2+\Lambda_3+4)(\Lambda_1+\Lambda_2+\Lambda_3+3)(2\Lambda_1+2\Lambda_2+\Lambda_3+5)
\eea
and 
\bea
C(\Lambda) &=& \Lambda_1^2 + 2 \Lambda_2^2 + \frac{3}{4} \Lambda_3^2 + 2 \Lambda_1\Lambda_2 + \Lambda_1\Lambda_3 + 2\Lambda_2\Lambda_3 \cr
&&+ 5 \Lambda_1 + 8 \Lambda_2 + \frac{9}{2} \Lambda_3
\eea
For the first few representations we find the following dimensions,
\bea
\dim(1,0,0) &=& 7\cr
\dim(0,1,0) &=& 21\cr
\dim(0,0,1) &=& 8
\eea
These correspond to the vector, the antisymmetric rank-2 tensor and the spinor representations of $SO(7)$ respectively. 

The Casimir invariant for $\Lambda = (n,0,0)$ becomes
\bea
C{(n,0,0)} &=& n^2 + 5 n
\eea
Now let us compare this Casimir invariant with the eigenvalues of the Laplacian acting on scalar harmonics. We use the method where we embed $S^D$ in $\mb{R}^{D+1}$. Spherical harmonics of rank $n$ are given by 
\bea
\t Y_n &=& C_{i_1\cdots i_n} x^{i_1} \cdots x^{i_n}
\eea
where $C_{i_1\cdots i_n}$ are symmetric and traceless. In polar coordinates in $\mb{R}^{D+1}$ with the metric 
\bea
ds^2 = dx^i dx^i = dr^2 + r^2 g_{MN} d\theta^M d\theta^N
\eea
we get
\bea
\t Y_n &=& r^n \t Y_n(\theta^M)
\eea
We note that on $\mb{R}^7$ we have
\bea
\partial_i^2 \t Y_n &=& 0
\eea
Expanding this relation in spherical coordinates, we get
\bea
\triangle Y_n &=& \frac{1}{r^2} \(n^2 + (D-1) n\) Y_n
\eea
where $\triangle$ denotes the Laplacian on $S^6$. Here $D = 6$ and therefore 
\bea
\triangle Y_n &=& \frac{1}{r^2} \(n^2 + 5 n\) Y_n
\eea
Thus we see that the Casimir invariant on the representation $(n,0,0)$ matches the eigenvalue of the scalar Laplacian on $S^6$ of unit radius.

\subsubsection*{Scalar harmonics}
We find that
\bea
\dim (n,0,0) &=& \frac{1}{120} (n+1)(n+2)(n+3)(n+4)(2n+5)
\eea
The first few scalar harmonics are
\bea
Y_0 &=& C\cr
Y_1 &=& C_{i_1} x^{i_1}\cr
Y_2 &=& \(C_{i_1 i_2} - \frac{1}{7} \delta_{i_1 i_2} C^j_j\) x^{i_1} x^{i_2}\cr
Y_3 &=& \(C_{i_1 i_2 i_3} - \delta_{(i_1 i_2} C^j_{|j|i_3)}\) x^{i_1} x^{i_2} x^{i_3}
\eea
The corresponding dimensions are
\bea
\dim(1,0,0) &=& 7\cr
\dim(2,0,0) &=& 27\cr
\dim(3,0,0) &=& 77
\eea
where $27 = \frac{7\cdot 8}{2} - 1$ and $77 = \frac{7\cdot 8\cdot 9}{1\cdot 2\cdot 3} - 7$.

\subsubsection*{Vector harmonics}
When we take the tensor product of a vector with a scalar harmonics, we get a decomposition that contains the vector harmonics,
\ben
(n,0,0) \otimes (1,0,0) &=& (n+1,0,0) \oplus (n-1,0,0) \oplus (n-1,1,0)\label{vec}
\een
The vector harmonics have the dimension
\bea
\dim(n-1,1,0) &=& \frac{1}{24} n(n+2)(n+3)(n+5)(2n+5)
\eea
and the Casimir invariant
\bea
C(n-1,1,0) &=& n^2 + 5 n + 4
\eea

\subsubsection*{Two-form harmonics} 
When we take the tensor product of a two-form with the scalar harmonics, we find the two-form harmonics plus some other representations,
\bea
(0,1,0) \otimes (n,0,0) &=& (n,1,0) \oplus (n-1,0,2)\oplus(n,0,0)\oplus(n-2,1,0)
\eea
The two-form harmonics are $(n-1,0,2)$ whose Casimir invariant and dimension are given by 
\bea
C(n-1,0,2) &=& n^2 + 5 n + 6\cr
\dim(n-1,0,2) &=& \frac{1}{12}n(n+1)(n+4)(n+5)(2n+5)
\eea

\subsubsection*{Spinor harmonics}
We take the product of a spinor with the scalar harmonics and get spinor harmonics as
\bea
(0,0,1)\otimes(n,0,0)&=&(n,0,1)\oplus(n-1,0,1)
\eea

The eigenvalues for $p$-form Laplacians on spheres has been obtained before and in full generality in \cite{AY}. 
In Theorem 4.2 of this reference there is a list of eigenvalues for spheres of even and odd dimensions. If we specialize to $S^6$, then we can read off from this list the following eigenvalues. For the representation $(n,0,0)$ corresponding to $0$-form harmonics, we have $\lambda_n = n(n+5)$. For the representation $(n,1,0)$ corresponding to the 1-form harmonics we have $\lambda_n = (n+2)(n+5)$. For the representation $(n,0,2)$ corresponding to $2$-form harmonics we have $\lambda_n = (n+1)(n+4)$. These eigenvalues as well as the corresponding representations are in precise agreement with our results. Furthermore Proposition 2.3 in this reference says that the Casimir invariant equals the eigenvalues of the Laplacian.

\subsection{Branching rules under $SO(7) \rightarrow SO(6)$}
The representations for spherical harmonics of $SO(6)$ can be found in \cite{Bak:2016vpi}. The dimension of an irreducible representation of $SO(6)$ is given by\footnote{To distinguish representations of various Lie groups, we attach a superscript on the Dynkin labels.}
\bea
\dim (\Lambda_1,\Lambda_2,\Lambda_3)^{SO(6)} =& \frac{1}{12}&(\Lambda_1+1)(\Lambda_2+1)(\Lambda_3+1)\cr
&&(\Lambda_1+\Lambda_2+2)(\Lambda_1+\Lambda_3+2)\cr
&&(\Lambda_1+\Lambda_2+\Lambda_3+3)
\eea
The first few representations have the dimensions
\bea
\dim (1,0,0)^{SO(6)} &=& 6\cr
\dim (0,1,0)^{SO(6)} &=& 4\cr
\dim (0,0,1)^{SO(6)} &=& 4
\eea
corresponding to the vector and two Weyl spinor representations of $SO(6)$. The spherical harmonics are
\bea
s &=& (n,0,0)^{SO(6)}\cr
v &=& (n-1,1,1)^{SO(6)}\cr
b &=& (n-1,2,0)^{SO(6)} \oplus (n-1,0,2)^{SO(6)}\cr
f &=& (n,1,0)^{SO(6)} \oplus (n,0,1)^{SO(6)}
\eea
for scalar, vector, two-form and fermion harmonics. 

We have the branching rules 
\bea
(n,0,0)^{SO(7)} &\rightarrow & \bigoplus_{k=0}^n (k,0,0)^{SO(6)}\cr
(n-1,1,0)^{SO(7)} &\rightarrow & \bigoplus_{k=1}^n (k,0,0)^{SO(6)} \oplus \bigoplus_{k=0}^{n-1} (k,1,1)^{SO(6)}\cr
(n-1,0,2)^{SO(7)} &\rightarrow & \bigoplus_{k=0}^n \((k-1,2,0)^{SO(6)} \oplus (k-1,0,0)^{SO(6)} \oplus (k-1,1,1)^{SO(6)}\)\cr
(n,0,1)^{SO(7)} &\rightarrow & \bigoplus_{k=0}^n (k,1,0)^{SO(6)} \oplus (k,0,1)^{SO(6)}
\eea

\subsection{Braching rules under $SU(4) \rightarrow SU(3) \times U(1)_H$}
The branching rules under $SU(4) \rightarrow SU(3) \times U(1)_H$ were obtained in \cite{Bak:2016vpi} and we use the same notation here. The Dynkin labels for $SU(4) = SO(6)$ are related as
\bea
(\Lambda_1,\Lambda_2,\Lambda_3)^{SO(6)} &=& (\Lambda_2,\Lambda_1,\Lambda_3)^{SU(4)}
\eea
We have the branching rules
\bea
(0,k,0)^{SU(4)} &\rightarrow & \bigoplus_{p=0}^k (p,k-p)_{[0]}\cr
(1,k,1)^{SU(4)} &\rightarrow & \bigoplus_{p=0}^k \((p,k-p)_{[0]} \oplus ((p+1,k-p)_{[-3]} \oplus (p+1,k+1-p)_{[0]} \oplus (p,k+1-p)_{[3]}\)\cr
(2,k-1,0)^{SU(4)} &\rightarrow & \bigoplus_{p=0}^{k-1} \((p,k-p-1)_{[-3]} \oplus (p,k-p)_{[0]} \oplus (p,k-p+1)_{[-3]}\)\cr
(1,k-1,1)^{SU(4)} &\rightarrow & \bigoplus_{p=0}^{k-1} \((p,k-p-1)_{[0]} \oplus (p+1,k-p-1)_{[-3]} \oplus (p+1,k-p)_{[0]} \oplus (p,k-p)_{[3]}\)\cr
(1,k,0)^{SU(4)} &\rightarrow & \bigoplus_{p=0}^k \((p,k-p)_{[-3/2]}\oplus (p,k-p+1)_{[3/2]}\)
\eea
A subscript in square brackets attached to an $SU(3)$ representation as $(m,n)_{[q]}$ gives the $U(1)_H$ charge $Q$ as follows,
\bea
Q &=& m - n + q
\eea

\section{Geometry of $S^6$}
The volume of $S^6$ of unit radius is 
\bea
\Vol(S^6) &=& \frac{16\pi^3}{15}
\eea
The first term in the heat kernel of a scalar field is therefore
\bea
\frac{1}{(4\pi t)^3} \Vol(S^6) &=& \frac{1}{60t^3}
\eea
We define the Riemann tensor, the Ricci tensor and the scalar curvature as 
\bea
R^{\lambda}{}_{\mu\nu\rho} &=& \partial_{\rho} \Gamma^{\lambda}_{\mu\nu} - \partial_{\nu} \Gamma^{\lambda}_{\mu\rho} + \Gamma^{\tau}_{\mu\nu} \Gamma^{\lambda}_{\tau\rho} - \Gamma^{\tau}_{\mu\rho} \Gamma^{\lambda}_{\tau\nu}\cr
R_{\mu\nu} &=& R^{\lambda}{}_{\mu\nu\lambda}\cr
R &=& R^{\mu}_{\mu}
\eea
On S$^6$ of radius $r$, we have
\bea
R_{\lambda\mu\nu\rho} &=&  g_{\lambda\rho} g_{\mu\nu} - g_{\lambda\nu} g_{\mu\rho}\cr
R_{\mu\nu} &=& 5 g_{\mu\nu}\cr
R &=& 30
\eea 
We have the cubic curvature invariants
\ben
L_1 &=& R^3\cr
L_2 &=& R R_{ij}^2\cr
L_3 &=& R R_{ijkl}^2\cr
K_1 &=& R_{ij} R_{jk} R_{ki}\cr
K_2 &=& R_{ij} R_{kl} R_{ikjl}\cr
K_3 &=& R_{ij} R_{iabc} R_{jabc}\cr
K_4 &=& R_{ijab} R_{ijmn} R_{abmn}\cr
K_5 &=& R_{ijkl} R_{iakb} R_{jalb}\label{curv}
\een
whose explicit values on $S^6$ become
\bea
L_1 &=& 27000\cr
L_2 &=& 4500\cr
L_3 &=& 1800\cr
K_1 &=& 750\cr
K_2 &=& -750\cr
K_3 &=& 300\cr
K_4 &=& -120\cr
K_5 &=& -120
\eea

Let us now view $S^6$ as a $S^5$-bundle over the interval $\theta \in [0,\pi]$. We write the metric on $S^6$ as
\bea
ds^2_{S^6} &=& r^2 \(d\theta^2 + \sin^2\theta ds^2_{S^5}\)
\eea
where we subsequently write the metric on unit $S^5$ as
\bea
ds^2_{S^5} &=& \(d\tau + \kappa\)^2 + ds_{CP^2}^2
\eea
The metric on $CP^2$ is written as
\bea
ds^2_{CP^2} &=& d\chi^2 + \frac{1}{4} \sin^2 \chi \(\sigma_1^2 + \sigma_2^2 + \cos^2 \chi \sigma_3^2\)
\eea
The volume of $CP^2$ is computed as
\bea
\int_0^{\pi/2} d\chi \int_0^{4\pi} d\psi \int_0^{\pi} d\theta \int_0^{2\pi} d\varphi \frac{1}{8} \sin^3 \chi \cos\chi \sin\theta &=& \frac{\pi^2}{2}
\eea
and the volume of the unit $S^5$ is 
\bea
\Vol(S^5) = \Vol(fiber) \Vol(CP^2) = 2 \pi \frac{\pi^2}{2} = \pi^3
\eea
Now we view $S^6$ as a singular fibration with fiber being the Hopf fiber of $S^5$. The base-manifold $M_5$ has conical singularities at $\theta=0$ and $\theta=\pi$ and is otherwise a smooth five-manifold. The metric is
\bea
ds^2_{M_5} &=& r^2 \(d\theta^2 + \sin^2\theta ds_{CP^2}^2\)
\eea
The volume is computed as
\bea
\Vol(M_5) = r^5 \int_0^{\pi} d\theta \sin^4 \theta \frac{\pi^2}{2} = \frac{3 \pi^3 r^5}{16}
\eea
The first term in the heat kernel is expected to be
\bea
\frac{1}{(4\pi t)^{5/2}} \Vol(M_5) = \frac{3\sqrt{\pi}}{512}
\eea

When we reduce along a singular circle fiber, singularities may arise at points where the fiber shrinks to zero size. An example can be found in two dimensions. Consider $S^2$ with metric in polar coordinates
\bea
ds^2 &=& \sin^2\theta d\varphi^2 + d\theta^2 
\eea
Here $\varphi \sim \varphi + 2\pi$ parametrizes a circle fiber with radius $r(\theta) = \sin\theta$ that becomes zero at $\theta = 0$ and $\theta = \pi$. Reducing along the fiber, we get a one-manifold with the metric 
\bea
ds^2 &=& d\theta^2
\eea
The one-manifold is smooth everywhere, except at the endpoints of the interval.

We believe that our $M_5$ that we obtain from reducing $S^6$ along a singular fiber, is smooth everywhere. The metric is 
\bea
ds^2_{M_5} &=& d\theta^2 + \sin^2\theta \(d\chi^2 + \frac{1}{4} \sin^2 \chi \(\sigma_1^2+\sigma_2^2+\cos^2 \chi \sigma_3^2\)\)
\eea
Possibly dangerous points would be at $\theta=0$ and $\theta=\pi$ which are where the circle fiber has zero length. Let us examine the vicinity of $\theta=0$. For small $\theta$, and small $\chi$, the metric is
\bea
ds^2_{M_5} &=& d\theta^2 + \theta^2 \(d\chi^2 + \frac{1}{4} \chi^2 \(\sigma_1^2+\sigma_2^2+\sigma_3^2\)\) + \O(\chi^3,\theta^3)
\eea
Of course there is no singularity at $\chi=0$ since we know that $CP^2$ is smooth, and indeed the metric 
\bea
d\chi^2 + \frac{1}{4} \chi^2 \(\sigma_1^2+\sigma_2^2+\sigma_3^2\) = d\chi^2 + \chi^2 ds_{S^3_{unit}}^2
\eea
corresponds to a foliation of $R^4$ with three-spheres of radii $\chi\geq 0$. Also then there is no singularity at $\theta =0$ as we may think on the metric locally around that point as as a foliation of ${R}^5$ by four-spheres. So we conclude that the local geometry around $\theta=0$ and $\chi = 0$ is given by ${R}^5$, so there is no conical singularity at that point. As we move away from $\chi =0$ there does arise a conical singularity at $\theta=0$ and $\theta=\pi$ for each submanifold defined by a constant value of $\chi$. However, those singularities arise by the choice of a singular submanifold inside the bigger smooth manifold $M_5$ and so they are not real singularities.

\section{Hadamard-Minakshisundaram-DeWitt-Seeley coefficients}\label{Tseytlin}
We follow notations and conventions of \cite{Vassilevich:2003xt}. The heat kernel expansion 
\bea
K(t) &=& \frac{1}{(4\pi)^3} \(\frac{a_0}{t^3}+\frac{a_2}{t^2} + \frac{a_4}{t} + a_6 + \O(t)\)
\eea
starts with the HMDS coefficient
\bea
a_0 &=& \int d^6 x \sqrt{g} \tr(1)
\eea
where the trace counts the number of components $d$ of the field. For $S^6$ we have 
\bea
\frac{1}{(4\pi)^3} \frac{a_0}{t^3} &=& \frac{d}{60 t^3}
\eea
so we may write the heat kernel expansion in the form
\bea
K(t) &=& \frac{d}{60} \(\frac{1}{t^3} + \frac{\t a_2}{t^2} + \frac{\t a_4}{t} + \t a_6\) + \O(t)
\eea  
where
\bea
\t a_2 &=& a_2/a_0\cr
\t a_4 &=& a_4/a_0\cr
\t a_6 &=& a_6/a_0
\eea

\subsection{Massless scalar ghost}
We begin by the simplest possible situation of one massless scalar on $S^6$. The HMDS coefficients are given by \cite{Gilkey:1975iq}
\bea
a_0 &=& \int d^6 x \sqrt{g}\cr
a_2 &=& \frac{1}{6}\int d^6 x \sqrt{g} R\cr
a_4 &=& \frac{1}{360} \int d^6 x \sqrt{g} \(5R^2 - 2 R_{ij}^2 + 2 R_{ijkl}^2\)\cr
a_6 &=& \frac{1}{5040} \int d^6 x \sqrt{g} \Bigg(\frac{35}{9} L_1 - \frac{14}{3} L_2 + \frac{14}{3} L_3\cr
&& - \frac{208}{9} K_1 - \frac{64}{3} K_2 - \frac{16}{3} K_3 - \frac{44}{9} K_4 - \frac{80}{9} K_5\Bigg)
\eea
For other fields, or for other manifolds, there will be more terms appearing in these coefficients, but for one massless scalar on $S^6$, this is all there is. If we compute these coefficients for $S^6$, then we get
\bea
\t a_2 &=& 5\cr
\t a_4 &=& 12\cr
\t a_6 &=& \frac{1139}{63} 
\eea
where 
\bea
\frac{a_0}{(4\pi)^3} &=& \frac{1}{60}
\eea

\subsection{Conformally coupled scalar} 
We now consider turning on a conformal mass term, which amounts to turning on $E$ that is shifting the Laplacian. We then get and extra contibution \cite{Gilkey:1975iq}. Using the notation in \cite{Vassilevich:2003xt}, this extra contribution reads
\ben
a_6^E &=& \frac{1}{360}\int d^6 x \sqrt{g} \(60 E^3 + 30 E^2 R + 5 E R^2 - 2 E R_{ij}^2 + 2 E R_{ijkl}^2\)\label{a}
\een
We have 
\bea
E &=& - \frac{1}{5}R
\eea
By plugging this into (\ref{a}), we get
\bea
a_6^E &=& \frac{1}{5.7!} \int d^6 x \sqrt{g}\(- \frac{98}{5}R^3 + 28 R R_{ij}^2 - 28 R R_{ijkl}^2\)
\eea
Evaluating this on $S^6$, we get
\bea
a_6^E &=& - 18 \int d^6 x \sqrt{g}
\eea
Then adding the massless contribution, we get in total
\bea
a_6 = \(\frac{1139}{63} - 18\) \int d^6 x \sqrt{g} = \frac{5}{63} \int d^6 x \sqrt{g}
\eea
In other words 
\bea
\frac{1}{60} \t a_6 &=& \frac{1}{750}
\eea

\subsection{Vector ghost} 
Following the conventions and notations in \cite{Vassilevich:2003xt}, we have
\bea
[D_{\mu},D_{\nu}] v_i &=& - R^m{}_{ijk} v_k
\eea
We define 
\bea
D_{\mu} v_{ij} &=& \partial_{\mu} v_{ij} + [\omega_{\mu},v]_{ij}\cr
[D_{\mu},D_{\nu}] v_{ij} &=& [\Omega_{\mu\nu},v]_{ij}\cr
\eea
and there from it follows that 
\bea
\(\Omega_{\mu\nu}\)_{ij} &=& - R_{\mu\nu ij}
\eea
We define
\bea
D &=& - D^{\mu}D_{\mu} - E
\eea
while we have 
\bea
\triangle^{(1)} v_i &=& - D^{\mu}D_{\mu} v_i + R_{ij} v_j
\eea
from which we see that 
\bea
E_{ij} &=& - R_{ij}
\eea

The additional terms that we now need to evaluate are
\bea
\delta a_6 &=& \frac{1}{360}\tr\Bigg(-12 \Omega_{ij} \Omega_{jk} \Omega_{ki} - 6 R_{ijkl} \Omega_{ij} \Omega_{kl} + 4 R_{ij} \Omega_{im} \Omega_{mj} + 5 R \Omega_{ij}^2 + 30 E \Omega_{ij}^2\cr
&& + 60 E^3 + 30 E^2 R + 5 E R^2 - 2 E R_{ij}^2 + 2 E R_{ijkl}^2\Bigg)
\eea
Let us begin by the last line, and substitute $E_{ij} = - R_{ij}$. Then we get
\bea
&& \frac{1}{360}\Bigg( - 60 R_{ij} R_{jk} R_{ki} + 30 R R_{ij}^2 - 5 R^3 + 2 R R_{ij}^2 - 2 R R_{ijkl}^2\Bigg)\cr
&=& \frac{1}{360} \(- 60  K_1 + 32 L_2 - 5 L_1 - 2 L_3\)
\eea
Next, the first line is 
\bea
&& 12 R_{ijab} R_{jkbc} R_{kica} - 6 R_{ijkl} R_{ijab} R_{klba} + 4 R_{ij} R_{lmab} R_{mjba} + 5 R R_{ijab} R_{ijba} - 30 R_{ab} R_{ijbc} R_{ijca}\cr
&=& 12 R_{ijab} R_{jkcb} R_{ikca} + 6 R_{ijkl} R_{ijab} R_{klab} + 4 R_{ij} R_{imab} R_{jmab} - 5 R R_{ijab} R_{ijab} + 30 R_{ab} R_{ijcb} R_{ijca}\cr
&=& 12 K_5 + 6 K_4 + 4 K_3 - 5 L_3 + 30 K_3\cr
&=& - 5 L_3 + 34 K_3 + 6 K_4 + 12 K_5 
\eea
Adding the two, we get
\bea
&& - 5 L_1 + 32 L_2 - 7 L_3 - 60 K_1 + 34 K_3 + 6 K_4 + 12 K_5\cr
&=& 5 A_{10} - 32 A_{11} + 7 A_{12} + 60 A_{13} - 34 A_{15} + 6 A_{16} + 12 A_{17}
\eea
By noting that $K_1+K_2=0$ or $A_{13} - A_{14} = 0$ on $S^6$, this result is identical to 
\bea
5 A_{10} - 32 A_{11} + 7 A_{12} + 52 A_{13} + 8 A_{14} - 34 A_{15} + 6 A_{16} + 12 A_{17}
\eea
which is in agreement with \cite{Bastianelli:2000hi}. Inserting the values, we get
\bea
\delta a_6 &=& - \frac{338}{3}
\eea
and in total we get
\bea
a_6 = 6 \cdot \frac{1139}{63} - \frac{338}{3} = - \frac{88}{21} = \frac{8}{7} \cdot \(-\frac{11}{3}\)
\eea
Thus we reproduce the result in \cite{Bastianelli:2000hi}.

\section{Supersymmetric cancelation}\label{SUSY}
The partition functions for a two-form ($B$), conformally coupled scalar ($S$), and fermions ($F$) are given by 
\bea
Z_B &=& \prod_{n=0}^{\infty} \frac{(n^2 + 5 n + 4)^{d_n^V/2}}{(n^2 + 5 n +6)^{d_n^T/2}(n^2 + 7 n + 6)^{d_{n+1}^{S_0}/2}}\cr
Z_S &=& \prod_{n=0}^{\infty}\frac{1}{(n^2+5n+6)^{\frac{d_n^S}{2}}}\cr
Z_F &=& \prod_{n=0}^{\infty} \(n+3\)^{2 f_n} 
\eea
The M5 brane partition function is
\bea
Z_{(2,0)} = Z_B^{1/2} Z_S^5 Z_F = \prod_{n=0}^{\infty} \frac{(n^2 + 5 n + 4)^{d_n^V/4}\(n+3\)^{2 d_n^F} }{(n^2 + 5 n +6)^{d_n^T/4+5d_n^S/2}(n^2 + 7 n + 6)^{d_{n+1}^{S_0}/4}}
\eea
By noting that 
\bea
n^2 + 5n + 4 &=& (n+4)(n+1)\cr
n^2 + 5n + 6 &=& (n+3)(n+2)\cr
n^2 + 7n + 6 &=& (n+6)(n+1)
\eea
we may, at least naively, shift the arguments $n$ so that we get
\bea
Z_{(2,0)} = \prod_{n=?}^{\infty} n^{2 d_{n-3}^F + \frac{d_{n-4}^V}{4} + \frac{d_{n-1}^V}{4} - \frac{d_{n-3}^T}{4} - \frac{5 d_{n-3}^S}{2} - \frac{d_{n-2}^T}{4} - \frac{5 d_{n-2}^S}{2} - \frac{d_{n-5}^S}{4} - \frac{d_n^S}{4}}
\eea
Of course this kind of manipulation for an infinite divergent product is not legitimate, and so there is no point in trying to make precise what happens at the lower summation point as we have indicated by $n=?$. Nevertheless, we think that it is interesting to compute the exponent, which simplifies drastically. We get
\bea
Z_{(2,0)} = \prod_{n=?}^{\infty} n^{-2n}
\eea
While the naive expectation might have been that we would get a fifth order polynomial in the exponent, we instead get a first order polynomial, 
\bea
P_{(2,0)}(n) &=& -2n
\eea
This shows that there is a huge cancelation of modes thanks to fermi-bose cancelations, which is typical behavior of a superconformal index.

We can repeat the same computation for the $(1,0)$ tensor multiplet, consisting of one scalar, two $SU(2)$-Majorana-Weyl fermions, and a selfdual two-form. The partition function is
\bea
Z_{(1,0)} = Z_B^{1/2} Z_S (Z_F)^{1/2} = \prod_{n=0}^{\infty} \frac{(n^2 + 5 n + 4)^{d_n^V/4}\(n+3\)^{d_n^F}}{(n^2 + 5 n +6)^{d_n^T/4+d_n^S/2}(n^2 + 7 n + 6)^{d_{n+1}^{S_0}/4}}
\eea
Again shifting $n$, we bring this into the form
\bea
Z_{(1,0)} = \prod_{n=?}^{\infty} n^{d_{n-3}^F + \frac{d_{n-4}^V}{4} + \frac{d_{n-1}^V}{4} - \frac{d_{n-3}^T}{4} - \frac{d_{n-3}^S}{2} - \frac{d_{n-2}^T}{4} - \frac{d_{n-2}^S}{2} - \frac{d_{n-5}^S}{4} - \frac{d_n^S}{4}}
\eea
Again the exponent simplifies, and we get
\bea
Z_{(1,0)} = \prod_{n=?}^{\infty} n^{\frac{n^3}{3} - \frac{7 n}{3}}
\eea
With less amount of supersymmetry, there is now less cancelation of modes and we end up with the cubic polynomial, 
\bea
P_{(1,0)}(n) &=& \frac{n^3}{3} - \frac{7 n}{3}
\eea

It would be very interesting if one could get these polynomials $P_{(2,0)}(n)$ and $P_{(1,0)}(n)$ from an index theorem and if one can find the corresponding nonabelian generalizations.

\end{document}